\documentclass[runningheads]{llncs}
\usepackage{graphicx}
\begin{document}
\graphicspath{{figures/}}
\title{Blockchain-based PKI for Crowdsourced IoT Sensor Information}
%
%
\author{Guilherme Pinto\inst{1} \and
Jo\~ao Pedro Dias\inst{2} \and
Hugo Sereno Ferreira\inst{2}}
\authorrunning{Guilherme Pinto et al.}
%
\institute{Faculty of Engineering, University of Porto, Portugal \\ \email{up201305803@fe.up.pt} \and INESC TEC and Department of Informatics Engineering, Faculty of Engineering, University of Porto, Portugal \\ \email{[jpmdias,hugosf]@fe.up.pt}}
\maketitle              
%
\begin{abstract}
The Internet of Things is progressively getting broader, evol-ving its scope while creating new markets and adding more to the existing ones. However, both generation and analysis of large amounts of data, which are integral to this concept, may require the proper protection and privacy-awareness of some sensitive information. In order to control the access to this data, allowing devices to verify the reliability of their own interactions with other endpoints of the network is a crucial step to ensure this required safeness. Through the implementation of a blockchain-based Public Key Infrastructure connected to the Keybase platform, it is possible to achieve a simple protocol that binds devices' public keys to their owner accounts, which are respectively supported by identity proofs. The records of this blockchain represent digital signatures performed by this Keybase users on their respective devices' public keys, claiming their ownership.
Resorting to this distributed and decentralized PKI, any device is able to autonomously verify the entity in control of a certain node of the network and prevent future interactions with unverified parties.

\end{abstract}
\section{Introduction} \label{sec:intro}

The Internet of Things concept gained popularity in the last couple of years. The convergence of the Internet with the RFID capabilities constituted, from the beginning, a powerful tool that provides great solutions for a wide variety of problems. With an interconnected network of smart devices and sensors, a large number of intelligent and autonomous services have been developed to improve personal, professional and organizational activities. \cite{Mio2012}

Together with the evolution of hardware components and the development of new communication protocols and technologies, several innovative projects are emerging. While smart houses are slowly becoming common, smart cities are projected for a near future and the integration with health-care, agriculture and wearables are gathering more attention.

According to a Gartner's report \cite{gartner_2017}, the consumer segment has settled itself as the largest supporter of the Internet of Things concept, with a total of 5.2 billion devices acquired in 2017, a number that is predicted to increase to the 7 billions by the end of 2018. Also the cross-industry seems to be progressively investing in the technology with smart-buildings taking the lead for their low-cost and highly interconnected devices.

As the number of users and devices grows, a larger amount of sensitive data is generated, reflecting the state of each supervised environment, objects or human beings' conditions. Given the low memory capacity of the majority of these devices, the data is required to flow through the network in order to be stored in specific nodes. At the end, the gathered information can be transmitted to application endpoints, where the data is analyzed and used for decision-making tasks or statistical studies on various subjects.

Besides confidentiality and privacy, trust became an important factor for any IoT system. The overall information shared through the network may, sometimes, require each smart device to properly identify the origins of the received packages as well as the recipients that it pretends to communicate with. Thus, defining a consistent identity management system, capable of satisfying the authenticity of each device towards the network, stands as a necessity to the evolution of trustworthy systems.

In this paper, it is presented an IoT solution that focuses on a blockchain implementation adapted to the Public Key Infrastructures roles on linking identities to public keys. In this case, it is pretended to link every device public key to a specific person and ensure that every action performed by a certain node of the network can be assigned to a proper entity.

The document begins by introducing the relevant concepts for this research, on Section 2, followed by the description of the problem to be addressed and which scenarios should be approached, in the third section. Further, in Sections 4 and 5, the implementation of the blockchain-based PKI is presented and the results of the experiments recreated are discussed, respectively. Ending the paper, the Section 7, concludes with some general observations over the research presented and which improvements could be done in the future.

\section{Background} \label{sec:sota}

\subsection{Blockchain} \label{sec:blockchain}

A Distributed Ledger Technology (DLT) is a collection of data, shared and synchronized across multiple individuals on a network, allowed to be geographically spread through distinct locations. It does not rely on a central administrator or centralized system for storage or management of data and it is supported on peer-to-peer networks, with consensus algorithms that ensure the replication of registries across the nodes. \cite{Deloitte2016}

Blockchain came up as an implementation of a distributed ledger. The term traces back to the Satoshi Nakamoto's original whitepaper from 2008, where he applied it as the core component of the digital currency named Bitcoin. \cite{Nakamoto2008} A blockchain is defined as a decentralized database, structured as a continuously growing list of ordered blocks, identified by a cryptographic signature. These blocks are linked in the chain by referring to the signature of the immediately previous record in the list. The blockchain contains an immutable collection of records of all the transactions processed, presenting a transparent, decentralized and secure solution for many applications. \cite{Jacobovitz2016}

\begin{figure}[h]
	\begin{center}
		\leavevmode
		\includegraphics[width=0.7\textwidth]{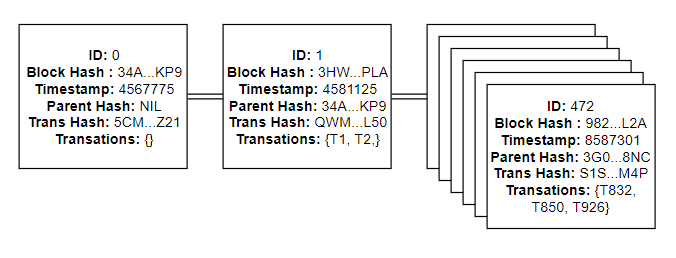}
		\caption{A representation of the relation between the blocks in a Blockchain.}
		\label{fig:blockchainSample}
	\end{center}
\end{figure}

The Figure \ref{fig:blockchainSample} illustrates how valid records are stored in a blockchain. The structure starts with a \textit{genesis block}, an initial entry that marks the beginning of the collection of information. Following it, every block in the blockchain should contain the information respective to a specific transaction. Each of these transactions must be digitally signed by the entity that is emitting it, constituting a \textit{block}. Upon a new block generation, the hash of the last block is retrieved and this new entry ends up referencing that previous record, becoming immediately linked to it. \cite{Samaniego2017}

It can be assumed that a blockchain acts as a state transaction system, where each state holds a snapshot of every transaction made until its creation. After the introduction of a new transaction, a new block is generated and a new snapshot is taken as a representation of this new state of the system. \cite{Dias2018}

Every blockchain can be inserted into one of the three categories that Vitalik Buterin summarizes in \cite{Vitalik2018}: public, consortium or fully-private. In the first category, anyone in the world can read, send transactions to be approved on the blockchain and even participate in the consensus process to determine which blocks will be validated and added to the chain. The consensus on consortium blockchains is controlled by a pre-selected set of participants, where a portion of them have to sign a block in order to validate it. When it comes to read permissions, these can be public, restricted to a certain target audience or even hybrid, with routes that define different levels of permission. The fully private blockchains are designed so that the write permissions are kept centralized to a specific authority. However, read permissions can also be public or restricted, like in the consortium blockchains.

\subsection{Public Key Infrastructure} \label{sec:pki}

Public key cryptography requires users to hold a key pair composed by a public and a private keys. However, it is important to assure that a certain pair of keys is linked to a specific entity. The Public Key Infrastructures can be interpreted as systems that properly manage these same public keys and provide a record to authenticate the link between them and their respective owners. Usually, these records are based on digital certificates that verify the ownership of a public key (and its corresponding pair, the private key) by some entity. Furthermore, it is expected from a PKI that it supports a set of functionalities comprising registration and update of public keys, as well as revocation or backup of certificates. \cite{technoPKI2018}

Generally, there are two approaches to Public Key Infrastructures. The most common one is the Certificate Authority-based (CA-based) PKI where the CA is considered a trusted third party and must be considered that way by every party involved in a transaction. The role of this Certificate Authority is to issue certificates that authenticate the link between a public key and its rightful owner. To trust in a Certificate Authority, every entity must accept a root certificate for that CA in its own collection. From this root element, it branches through a hierarchical certificate chain, where any certificate signed by a trusted certificate is consequently trusted. \cite{Axon}

Considering that most of the PKI's basic are directly related to PGP's concepts, one other convectional approach is based on the Web of Trust definition. Unlike the previously mentioned CA-based PKI, the trust becomes decentralized. In this case, a certain signature on a given public key is trusted by a user if it was already verified by a certain trusted party.

\subsection{Pretty Good Privacy} \label{sec:pgp}

Pretty Good Privacy is an encryption standard developed by Phil Zimmermann, in 1991, that provides cryptographic privacy and authentication for communications. It can be used for signing, verification, encryption and decryption of digital data such as emails, files, text or entire directories. \cite{Shaw2007}

\begin{figure}[h]
	\begin{center}
		\leavevmode
		\includegraphics[width=0.6\textwidth]{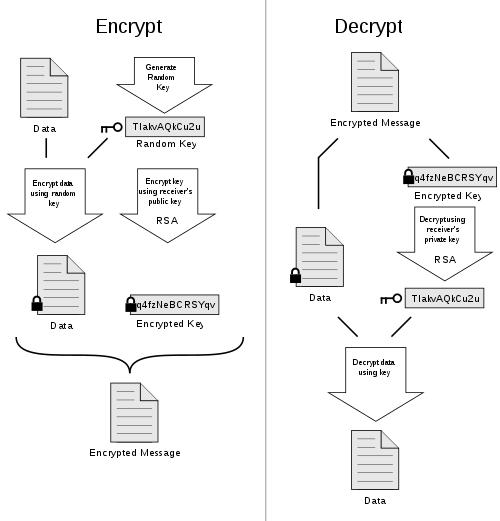}
		\caption{Encryption and Decryption procedures in the PGP standard. \cite{PGPWiki}}
		\label{fig:pgp}
	\end{center}
\end{figure}

PGP operations combine data compression with hashing and symmetrical key with public key cryptography to provide integrity and confidentiality of data. When information is exchanged between two actors, it is encrypted using a symmetric encryption algorithm. The symmetric key, known as \textit{session key}, is used only once for encryption and decryption, and is refreshed as a new random number in each cycle of communication. The data is sent to the receiver together with session key, encrypted with the receiver's public key. Every public key is, therefore, linked to a single user and will be utilized by other entities anytime they pretend to send him confidential information. Upon obtaining the data, a receiver must first decrypt the session key with is own private key, which is kept personal, and decrypt the message with the provided session key. An illustration on the same procedure is shown in the Figure \ref{fig:pgp}.

Authentication via digital signatures can be also achieved with PGP, thanks to hashing and public-key cryptography. In a similar process to the one previously described, upon creating a message, the sender generates an hash code (also known as a fingerprint of the data), and encrypts it using his own private key. The hash is now attached to the message and sent to the receiver. This time, the receiver must also create a fingerprint from the message and decrypt the received hash with the sender's public key. If the generated hash is compatible to the received one, the authentication is successful.

Currently, PGP is mainly adopted for emailing systems, but has also been implemented in digital signatures management, full encryption of memory partitions, directories and instant messaging session protection and, most recently, for the encryption and signature of HTTP requests and responses through server and client modules.

OpenPGP, which will be covered later on the document, is an open-source standard for the world, under the RFC4880, and currently on the Internet Standards Track. It presents several specifications on encryption and decryption operations that are followed by the majority of PGP applications.

\subsubsection{Web of Trust} \label{sec:wot} is commonly introduced in systems that implement the Pretty Good Privacy (PGP) standard, aiming to establish the authenticity of the link between a specific public key and the respective owner, through a decentralized trust model.

When some message is encrypted with the public key of a targeted receiver, is is important to know that the key belongs to the intended used. Assuming that impersonation is a reality in any vulnerable network, simply loading a public key from a public directory does not guarantee this association between a key and a real identity.

The binding refers to the relation between a pair of public and private keys and the identity of a specific person or organization. This relation is usually defined through verifications upon interactions with other entities. Lets assume that a user named Alice witnessed that Charlie is in the possession of a pair of PGP keys and signed his public key with her own private key, in order to vouch for him. If Charlie intend to email another user, called Bob, who doesn't know him, he might not trust Charlie right away. However, if Bob had previously verified Alice and signed her key, thus trusting her, then he can indirectly assume that Charlie is also trustworthy upon acknowledging that Alice signed his key. With this model, the more people sign each others keys, the shorter the trust paths between parties in a Web of Trust become. \cite{Fromknecht2014}

The PGP's Web of Trust concept existed for over 20 years. However, technically, it is difficult to implement and utilize, requiring personal verifications and becoming hard to know which trust level should be assigned on each verification. In theory, the concept proves to be successful, except for the con that implies that people are needed to validate a person's possession of a GPG public key before signing it.


\section{Problem}

With the number of identities in the IoT environment tending to grow, it becomes urgent for any platform to have the resources capable to manage them. According to \cite{Kaffel-BenAyed2017}, in order to implement a consistent Identity Management System for the Internet of Things, some properties like privacy, security, mobility and trustworthiness must be ensured.

In general, these "things" present a direct relationship with real people, concerning ownership or manufacturing, for instance. In IoT, these ownership and identity relationships with real entities present a substantial impact on the systems' identity processes like authentication and authorization as it must be required to determine, rigorously, who is the owner of a certain node on the network.

The main issues that this paper pretends to target is the centralized data collection of identities and management of devices, which is a method that doesn't scale in the context of IoT. In fact, managing billions of devices constantly exchanging messages between themselves in a smart and dynamic network, cannot be efficiently implemented in a centralized system. \cite{Moratoes2015}

Personal and private networks don't require the verification of the ownership from each node in the network. Usually, when a device is included in these types of networks, they should go through a strict process of configuration and installation due to the system's operations being closed to the external environment.

The attention for this research focuses on specific kinds of networks, considered public, where multiple entities may participate while being globally distributed, if necessary. In this case, with a more opened environment and free participation of unknown parties, it becomes necessary to determine the identity behind each node and assign the responsibility of its interactions to a specific person.

The following examples provide real implementations of networks that represent the type of cases this research pretends to support.

\begin{description}
	\item[AccuCast] is a service provided by AccuWeather, a global leader in weather information. This global service, launched in 2015, allowed an interactive network where iOS users can share their local weather updates through the AccUcast application. This idea was designed with the objective to help people around the world to make more informed decisions, providing a new level of localization and user interactivity in the weather forecasting process. \cite{AccUWeather2015}
	\item [Light Pollution Map]	consists on a global system of small devices called Sky Quality Meters (SQM) and Sky Quality Cameras (SQC) that provide a set of measurements to extract data about the consumption of energy as lightning in different locations of the globe. \cite{LightPolMap2018}
	\item [uRADMonitor] is a network composed by IoT devices equipped with sensors for environmental monitoring cities, offices and homes, spread in more then 40 countries to generate uniform and comparable environmental data to be used on the analysis of current global pollution. \cite{uRad2018}
\end{description}

In both cases, if participants decide to provide fake information, the services becomes untrustworthy and unreliable. However, if a system detects which device contributed with false information and the responsible entity gets identified, the system could apply proper punishments to this actor.

\section{Blockchain-Based PKI}

The goal of the research is to develop a distributed infrastructure capable of registering the devices admitted in a network into a safe and verifiable data structure. However, this same infrastructure must be supported by an external system for identity management, where every user is supported by proofs that authenticate himself and assign an absolute level of trust that link that user account to the person that it belongs to.

\subsection{High-Level Overview}

\subsubsection{Keybase} can be summed up as a collection of tools that establish a Web of Trust, associating their users' accounts to the most common social networks, such as Facebook, Twitter or Reddit. In order to achieve this, Keybase implements Pretty Good Privacy (PGP) policies, assigning keys to each account which will be used to support a set of proofs that link the same user to other external accounts from distinct networks. This is done through signed statements posted in accounts that a user wants to prove ownership of. These constitute a publicly verifiable collection of identity proofs that can be individually verified to ensure trust on the interactions established with each account.

Assuming this support for PGP encryption and identity management of users, Keybase holds a set of capabilities useful for the objective in study. The signature of artifacts with the keys associated to each account create a direct link between the respective user and his own digitall properties. Additionally, the publicly verifiable proofs that the platform administers are sufficient to ensure a truthful entity behind each user account on Keybase. However, considering the relevance of this platform for the system in mind, it becomes crucial to find a way to interact with the tools held by Keybase and integrate them into the solution to be developed.

For the infrastructure to be developed, Keybase supplies just the enough resources to support it. Two of the actions required from the encryption operations that PGP offers are the signature of data and the respective verification of these originated statements. In order to implement them, the following must be ensured: the private key of a Keybase account must be securely exported from the platform, so that the respective owner can use it to sign external data; in the other hand, Keybase must be able to provide a user's public data, with its corresponding public key and account information.

Keybase's Command Line provides a specific set of commands to extract both keys assigned to an account on the platform. This pair of PGP encryption keys must be previously generated by the owner of the account in order to continue the procedure. At this moment, the priority is to extract the private key associated the user and import it into the local GPG system. It can only be done if the user's session is set in the local Keybase application and the pass-phrase for the private key will also be prompted. Once the key is successfully retrieved, GPG allows the user to write it into a visible file in the computer. With this file, the owner can then carefully use it for digital signature purposes.

The system still requires the retrieval of the information associated to a specific user account, with special urgency for the public key, for signature verifications. In this case, the Keybase API holds the \textit{user/lookup} endpoint: a public API call that retrieves a profile's information given its username. This request answers with a complete structure of the profile that was queried. With it, a field containing the respective public key is presented. An application can easily ask for this public information and evaluate any signature after retrieving the proper public key.

\subsubsection{A PKI built on the Blockchain} constitutes the second component of the system. It aims to hold a set of digital signatures that link each device to a single Keybase account. It must be guaranteed that any party on a network of devices must always be owned by some entity. The blockchain will act as the collection of records that will connect both ends in order to assign each device action to a specific person or organization. 

In order to define how blockchain and PKI's can complement each other, it is convenient to expose their functions. While the blockchain supports the distribution of transactions and registration of the blocks in a secure and reliable way, the Public Key Infrastructure deals with the registration and revocation of digital certificates that are proof of the ownership of a public key by a specific identity. In this case, the blocks' structure must be adapted to the objective, displaying the necessary data to implement a protocol that easily verifies the data associated to both devices and Keybase users.

\begin{figure}[h]
	\begin{center}
		\leavevmode
		\includegraphics[width=0.6\textwidth]{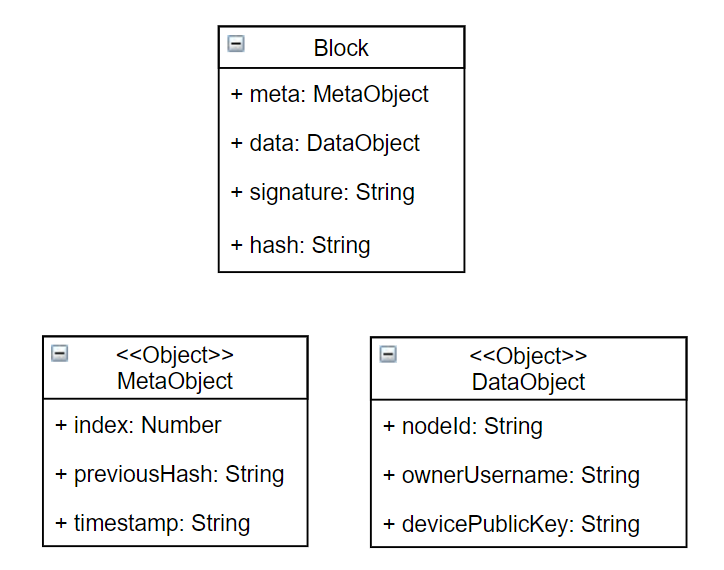}
		\caption{Block structure of the blockchain}
		\label{fig:block}
	\end{center}
\end{figure}

The figure \ref{fig:block} provides a representation of the data to be included in the registries of the blockchain. Each of the blocks will contain a meta data field, which is responsible to hold the position of the block in the chain, the hash of the immediately previous block and the timestamp, in order to verify the chronological order of the entries. The second field is an object with the relevant information on the device to be introduced to the system. It will contain the identification of the node, the Keybase username of the owner and the public key generated for the device. The username of the owner is an essential field to retrieve the identity information on Keybase, via the public endpoint of the API mentioned earlier. The signature property, as the name suggests, will provide a digital signature of the data object, produced with the private key of the owner, exported from Keybase. Finally, the hash field displays a fingerprint of the complete record, based on the three previously mentioned fields.

\subsection{The Protocol} \label{sec:protocol}

After the overview of the solution, it becomes important to clarify each step of the protocol to be developed. The following sections describe each of the steps that constitute the developed work and how they connect each of the components of the system to achieve the objective in mind.

\begin{figure}[h]
	\begin{center}
		\leavevmode
		\includegraphics[width=0.6\textwidth]{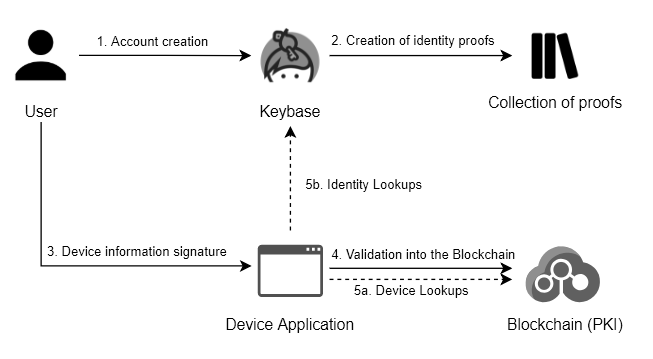}
		\caption{Abstraction of the solution's relevant components and how they should interact to ensure the linkage between users and devices.}
		\label{fig:diagram_relations}
	\end{center}
\end{figure}

\subsubsection{User Registration and Requirements:}

The initial stage of the procedure consists on creating a Keybase account. In fact, it is the chosen platform to handle the identity management of users or eventual organizations and, without digital records of entities, it is not possible to associate any kind of data to someone.

The first and second steps illustrated on the Figure \ref{fig:diagram_relations} are associated to this user registration. Every action that the actor is required to do can be executed in the Keybase's application, where he can insert his personal information and follow the registration tasks imposed by the platform. Keybase might even prompt every user to go through a proper protocol to link devices in order to strengthen the bind between user and the account.

In order to become suitable to this solution, every account in Keybase must be completed with a few proofs that the platform supports. As mentioned previously, these proofs are public and can be validated by anyone in order to vouch for the user behind the account. The first and most relevant proof for a user profile is the creation of a pair of PGP keys. This pair will provide means for the user to sign public statements that will prove his ownership of other social network accounts. These same statements also act as proofs on Keybase and are also a requirement in this protocol, as they are fundamental to guarantee the trust on a specific user account.

Unfortunately, the Keybase API doesn't provide any endpoint to submit a custom package of information to be digitally signed by the authenticated user. In order to gain control of these keys, Keybase enables any user to export any of his PGP keys through Command Line calls. 

To be able to execute the mentioned commands, it is demanded to the user not only to be locally signed into the Keybase application, but also to have the PGP keys generated in his own account, has referenced earlier. It is also required a local PGP application such as the GNU Privacy Guard software, to handle the extracted cryptographic keys. 

\subsubsection{Device Signature:} \label{device-sig}

The second stage of the protocol consists on composing a structure of information relating a certain device to its public key and Keybase owner, who should digitally sign it and submit the output as a new transaction to the blockchain, which is acting as a Public Key Infrastructure.

It is considered a total of three fields that the author assumes to be required in order to create the pretended association: the device unique identification number, the Keybase username of the owner and the public key of the device.

The user should only be required to upload the \textit{private key} extracted in the first phase together with the pass-phrase that unlocks it and the Keybase username. A new record is generated and the data is introduced to the blockchain as a new block that can be verified by any other device that pretends to interact with the signed equipment.

\subsubsection{Ownership Verification:}

When an interaction between two unknown devices occurs, it needs to be established a set of verifications by both sides that displays information about the entities involved in the communication. For a certain device to validate another device, it must first check on its owner's identity, verify the proofs that he presents and check whether they are trustworthy or not.

\begin{figure}[h]
	\begin{center}
		\leavevmode
		\includegraphics[width=0.6\textwidth]{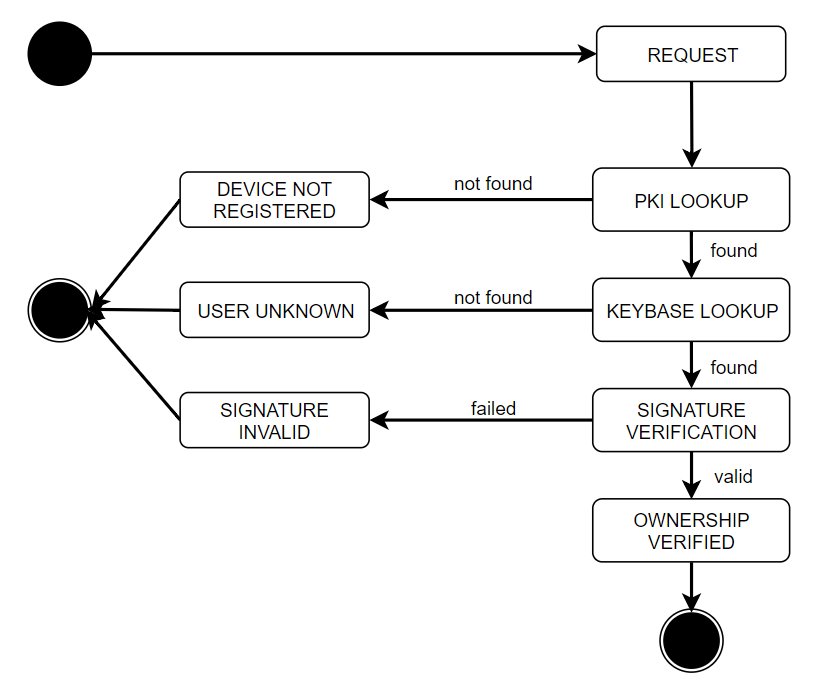}
		\caption{Representation of the states during the ownership verification process of devices, comprising a set of validation steps required to verify that a device was really acknowledged as a property of a properly identified Keybase user.}
		\label{fig:protocol3}
	\end{center}
\end{figure}

In the protocol, the verification can be done through look-ups on the blockchain and through queries to the Keybase API, as illustrated in the steps 5a and 5b of the Figure \ref{fig:diagram_relations}. Upon receiving data or a request from a strange device B, the device A must, first of all, verify for a registry in the blockchain that refers to B. This block will, as explored before, contain the signature of the information about this device and provide its respective public key and the username of the Keybase owner. Gathering this username, it is simple to query for the user on the public endpoint of the Keybase API. This call, if the username really exists, will provide a set of data concerning the Keybase account of the owner. Amongst this information is the public key of the user that can be applied to verify the block signature. If such is positively validated, the device A can then check on the number and type of proofs held by the account and decide whether or not to trust and continue to communicate with B. The diagram provided in the Figure \ref{fig:protocol3} illustrates the possible states on this third protocol phase.

\section{Experiments \& Observations}

In order to take conclusions on the implemented solution, few experimental use cases are required to explore the most common situations that the system could face during execution. With it, it is intended to put the solution to the test and observe how it reacts towards each of them. Also, this practical method provides the opportunity to identify weaknesses on the implementation and target a few goals for future development.

\subsubsection{Use Case \#1 - Device Signature Verification}

An interaction between two devices implies that a certain gadget is requesting to communicate with another device.

The implemented solution is simulating this request as a single \textit{ping} to the network, where a device notifies every other connected nodes with a simple and unencrypted message containing its \textit{nodeId}. Receiving this packet, a device can then lookup for the device that emitted the message and verify its ownership.

Upon retrieving the information of the owner of the sender device, the receiver can then proceed with the verification of the former entity. From this instance, it is possible to face one out of three situations: 

\begin{enumerate}
	\item The public key retrieved from the user's Keybase account doesn't validate the block signature;
	\item The public key retrieved from the user's Keybase account validates the block signature but the user doesn't provide enough identification proofs;
	\item The public key retrieved from the user's Keybase account validates the block signature and the number of associated proofs complies with the minimum level imposed by the protocol.
\end{enumerate}

The receiver is able to easily verify the signature of the block after collecting the public key and conclude whether or not the entity that signed the record really owns the appropriate secret key. If he doesn't, then it is assumed that he isn't the rightful owner of the sender device.

In the situations 2 and 3, the signature is successfully validated with the public key collected. However, depending on the configuration on the receiver device, it may require a higher or lower number of proofs in order to accept the communication with the requester. These proofs can be analyzed together with the information retrieved and the experiments proved that the implementation is able to respond correctly to these requirements.

\subsubsection{Use Case \#2 - Unreliable User Proofs}

It is possible to create multiple fake accounts on Facebook, Reddit or Twitter and sign statements from a Keybase account to claim ownership over them. The most certain is that this fake account in Keybase will not be followed by any other user.

However, the implemented system is not prepared to consider how many followers a certain account has for a very simple reason: there is no API call nor resources provided by Keybase that return the number of followers that a given account has. The support of these complementary proofs would guarantee and extra parameter to consider and judge more accurately the identity behind the devices, being sure that their public statements were verified by other real entities. If a certain device is receiving interactions from two other nodes of the network, it may discard legitimate entities and approve malicious ones due to lack of complementary user data.

\subsubsection{Use Case \#3 - Revocation of the user's PGP keys}

The third use case contemplates the possibility of a Keybase user revoking his own PGP keys and replace them by another generated pair, associating it into the account. The Keybase platform is prepared to handle these actions and propagate the necessary changes on the user's chain of proofs.

However, these actions of revocation and update of the PGP keys of a certain device's owner brings a negative impact to the implemented solution. As result of the experiments made, the association of a new pair of PGP keys to a certain device owner leads to unsuccessful identity checks on the Keybase platform. This happens because the protocol will look up to the user profile, gathering his current public key and verify the device signature with it. If this signature was created with the previous pair of keys, the result will be to consider it as a bad device signature and it would be impossible to associate the devices to a rightful owner.

\subsection{Results Evaluation} \label{evaluation}

Theoretically, blockchain's design provides security characteristics that are easily adapted to the Public Key Infrastructure concept, allowing for certificate transparency on signatures and revocations, a reliable collection of transactions and, due to its distributed and decentralized nature, allows the elimination of potential points-of-failure created by the adoption of Certificate Authorities in conventional PKI systems.

From the observed interactions with the conceived protocol, blockchain-based PKI's also show the potential to overcome the Web-of-Trust PKI's. While the WoT model requires a significant effort to produce a web capable of proving the trustworthiness of a node to a considerable portion of the network, blockchain-based PKI's don't require such an interconnected structure of authenticated entities.

The presented approach is built on top of a platform that already implements this Web of Trust definition through the PGP operations inherent to its functionalities. The creation of public and verifiable statements that support the identity of each user allow them to be validated by other real life entities and vouched through follower statements. Assuming the trust between a set of entities, the devices owned by them could also be able to interact with each other. Adopting these WoT characteristics from Keybase, the effort required to implement the Web-of-Trust PKI would be minimized and the result of this research could be different.

As denoted in the second use case, Keybase does not provide these follower statements neither a qualification of the trust between two given entities. Consequently, the intention to develop a WoT PKI becomes somehow more complex. However, Keybase allows anyone to access a public endpoint of the API to retrieve the ownership statements of external social networks accounts and digital assets, which became the source of trust in the blockchain-based PKI that is presented.

The implementation described during the previous chapters aims to transparently display a collection of signatures over the devices participating in a network. Resorting to this collection of signatures, securely appended to a blockchain that is distributed amongst the devices, they can check on the ownership of every machine they interact with and assign the actions of that device to a specific entity that claimed its ownership. The first use case proves that the protocol is successful and that the devices take the proper actions to prevent or allow the interaction with other unknown devices, based on the proofs provided by their owners on Keybase.

Considering the third use case, there are still a few operations that could be integrated and experimented, with special attention to the revocation of users' PGP keys in Keybase. The revocation and update of keys are common operations in Public Key Infrastructures and the protocol that was designed could be complemented with this functionality, providing more flexibility to the system and a more close approach to what PKI's should provide to their environments.

\section{Conclusions \& Further Work}

The research presented in this paper provides a different approach for the Internet of Things segment, adopting the distributed ledger technologies as bridge between the interaction amongst smart devices and the importance of social networks in today's society.

Considering the decentralized, distributed and immutable nature of the blockchain together with the purpose of Public Key Infrastructures to manage the ownership of multiple digital assets, it was explored an innovative and simple methodology to discover the entity behind a specific device. Resorting to a set of digital signatures and encryption operations, any action performed by a device can be assigned to its respective owner, through simple lookups into a PKI built on the blockchain and queries to the Keybase API, which provides the sufficient proofs to judge the identity in control of a node.

With this protocol, it is pretended to provide another feasible and secure implementation for the identity management of entities in the IoT systems, preventing malicious actors from anonymity and impersonation when introducing devices in the network that can't get their respective owner properly verified.

However, the implemented solution lacks on some functionalities and improvements that would benefit from the continuity of this research.

The most urgent task to implement would be the revocation and update of users' PGP Keys, from Keybase. This issue represents a fundamental functionality in every Public Key Infrastructure. Implementing it would prevent a device from being untraceable to the respective owner after the entity renewing his Keybase PGP keys.

The consensus algorithm implemented in the blockchain considers only the largest distributed version of the chain as the legitimate collection of records. This consensus, being computational simple and requiring low effort from the devices in order to process transactions, also provides weaknesses. If a certain device or group of devices with increased computational power works in order to introduce a longer and corrupted version of the blockchain, it may end up with tampered or fake registries.

One last aspect to look into would be the experiments on real devices and simulate the same tested use cases in a real and physical environment. This experiments would allow to explore which were the minimum requirements for a device to handle the protocol and how it would impact the efficiency of the system operations.


%
%
%
\bibliographystyle{splncs04}
\bibliography{mybibliography}

\end{document}